\documentclass[prb,twocolumn]{revtex4}
\usepackage{graphicx} 

\usepackage{amsmath,amsthm,amssymb,mathrsfs}
\usepackage{bm}

\begin{document}

\title{Periodic structure of memory function in spintronics reservoir with feedback current}

\author{Terufumi Yamaguchi${}^{1}$, Nozomi Akashi${}^{2}$, Sumito Tsunegi${}^{1}$, Hitoshi Kubota${}^{1}$, Kohei Nakajima${}^{2}$, and Tomohiro Taniguchi${}^{1}$
      }
 \affiliation{
 ${}^{1}$National Institute of Advanced Industrial Science and Technology (AIST), Spintronics Research Center, Tsukuba, Ibaraki 305-8568, Japan, \\
 ${}^{2}$Graduate School of Information Science and Technology, The University of Tokyo, Bunkyo-ku, 113-8656 Tokyo, Japan
 }

\date{\today} 
\begin{abstract}
{
The role of the feedback effect on physical reservoir computing is studied theoretically by solving the vortex-core dynamics in a nanostructured ferromagnet. 
Although the spin-transfer torque due to the feedback current makes the vortex dynamics complex, 
it is clarified that the feedback effect does not always contribute to the enhancement of the memory function in a physical reservoir. 
The memory function, characterized by the correlation coefficient between the input data and the dynamical response of the vortex core, 
becomes large when the delay time of the feedback current is not an integral multiple of the pulse width. 
On the other hand, the memory function remains small when the delay time is an integral multiple of the pulse width. 
As a result, a periodic behavior for the short-term memory capacity is observed with respect to the delay time, 
the phenomenon of which can be attributed to correlations between the virtual neurons via the feedback current.
}
\end{abstract}

 \maketitle



\section{Introduction}
\label{sec:Introduction}

Physical reservoir computing is a kind of recurrent neural network based on physical systems, such as optic lasers, soft materials, and quantum systems 
\cite{mandic01,maas02,jaeger04,verstraeten07,appeltant11,brunner13,nakajima15,fujii17,nakajima19,nakajima20}. 
The dynamical response of the physical system, called the reservoir, consisting of numerous virtual neurons with recurrent interactions 
reflects the history of time-dependent data inputted into the reservoir. 
Such a response can be applied to an energy-efficient brain-inspired computation calculating and classifying a time sequence of input data such as spoken language and movie. 
Therefore, physical reservoir computing has attracted much interest from various fields of science 
such as neural science, information science, and physics. 
Recently, it has been shown that a fine structure of ferromagnets on nanoscale is regarded as a physical reservoir 
\cite{torrejon17,furuta18,tsunegi18,bourianoff18,nakane18,markovic19,tsunegi19,riou19,nomura19}, 
and other computational schemes and related phenomena based on spintronics devices have also been proposed 
\cite{borders17,kudo17,romera18,mizrahi18,borders19,yoo20,daniels20}. 


An intriguing method for the physical reservoir computing is time-multiplexing method \cite{fujii17,torrejon17,tsunegi18}, 
which creates virtual neurons by dividing the dynamical response from the reservoir into several nodes, 
and thus, enables us to perform the reservoir computing by using small numbers of physical particles. 
A critical issue of this method is, however, low-memory functionality of the reservoir due to lack of richness in the dynamics. 
A fascinating approach to solving the issue is to use a feedback effect 
because it makes the dimension of the phase space infinite and causes highly nonlinear dynamics \cite{biswas18}, 
although adding a delay line eventually forces us to make the size of the whole device large. 
Recently, the role of the feedback effect on spintronic systems has been extensively studied \cite{khalsa15,tsunegi16SR,williame19,taniguchi19}. 
For example, a modulation of Hopf bifurcation in a vortex-type ferromagnet by the feedback of electric current 
was proposed theoretically \cite{khalsa15} and was confirmed experimentally \cite{tsunegi16SR}. 
Chaos in nanomagnets via feedback current was also theoretically predicted \cite{williame19,taniguchi19}. 
An enhancement of pattern-recognition rate in the presence of the feedback current was reported recently \cite{riou19}. 
It is, however, still not fully understood how the feedback effect relates to the memory function of the physical reservoir. 
More generally, in other words, the roles of network topology, complexity, and/or parameters on the performance of the reservoir computing remain unclear yet \cite{rodan10,kawai19,rodriguez19}.

In this work, a periodic behavior for the memory function with respect to the delay time is found in the spintronics reservoir having the feedback current. 
As a figure of merit, the short-term memory capacity is evaluated, which characterizes the number of the input data stored in a physical reservoir. 
The short-term memory capacity is maximized when the delay time $\tau$ of the feedback current is slightly longer than the pulse width $t_{\rm p}$ of the input data. 
The periodic reduction of the short-term memory capacity is also revealed when the delay time is an integral multiple of the pulse width, 
i.e., $\tau \simeq n t_{\rm p}$ ($n\in \mathbb{N}$). 
The periodic behavior of the memory function can be attributed to an interaction between the virtual neurons via the feedback effect. 


The paper is organized as follows. 
In Sec. \ref{sec:System description}, we give the description of the magnetic multilayer including a magnetic vortex and the equation of motion for the core center. 
In Sec. \ref{sec:Dynamical response to pulse current}, the dynamics of the magnetic vortex-core with respect to the pulse current is studied. 
In Sec. \ref{sec:Reservoir computing and periodic structure in memory capacity}, the short-term memory capacity is evaluated, where the periodic structure of the memory function is found. 
Section \ref{sec:Conclusion} summarizes the conclusions of this work. 


\section{System description}
\label{sec:System description}



\begin{figure}
\centerline{\includegraphics[width=1.0\columnwidth]{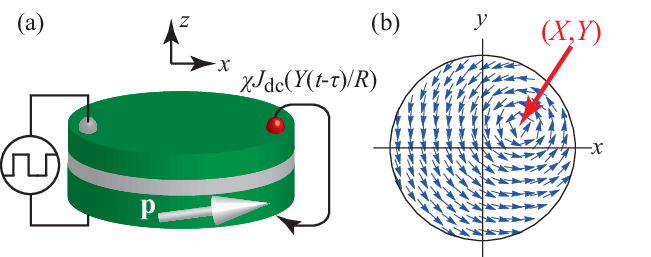}}
\caption{
        (a) Schematic picture of the magnetic trilayer. 
            The magnetization in the reference layer is $\mathbf{p}$. 
            The current density injected into the free layer consists of a bias and random binary pulse. 
            In addition, the feedback current ($\chi J_{\rm dc} [Y(t-\tau)/R]$) is also injected. 
        (b) A magnetic configuration in a vortex-type free layer. 
            The vortex-core center position is $\mathbf{X}=(X,Y,0)$. 
         \vspace{-3ex}}
\label{fig:fig1}
\end{figure}



The system investigated in this work is schematically shown in Fig. \ref{fig:fig1}(a). 
A magnetic tunnel junction (MTJ) consists of a vortex-type free layer and uniformly magnetized reference layer. 
The unit vector pointing in the magnetization direction of the reference layer is denoted as $\mathbf{p}$. 
The $z$ axis is perpendicular to the film plane, whereas $\mathbf{p}$ lies in the $xz$ plane as $\mathbf{p}=(p_{x},0,p_{z})$. 
Current density $J$ is injected into the free layer, 
where the positive current is defined as current flowing from the reference to free layer. 
The spin-transfer torque \cite{slonczewski96,berger96,slonczewski05} caused by the current compensates with the damping torque, and moves the vortex core from the disk center, as schematically shown in Fig. \ref{fig:fig1}(b). 
The dynamics of the vortex-core center is described by the Thiele equation for the core position $\mathbf{X}=(X,Y,0)$ given by \cite{thiele73,guslienko06PRL,guslienko06,khvalkovskiy09,guslienko11,dussaux12} 
\begin{equation}
\begin{split}
  &
  -G 
  \hat{\mathbf{z}}
  \times
  \dot{\mathbf{X}}
  -
  |D| 
  \left(
    1
    +
    \xi
    s^{2}
  \right)
  \dot{\mathbf{X}}
  -
  \frac{\partial W}{\partial \mathbf{X}}
\\
  &
  +
  a_{J} J 
  p_{z}
  \hat{\mathbf{z}}
  \times
  \mathbf{X}
  +
  c a_{J} J 
  R_{0} p_{x}
  \hat{\mathbf{x}}
  =
  \bm{0},
  \label{eq:Thiele}
\end{split}
\end{equation}
where $G=2\pi pc ML/\gamma$ and $D=-(2\pi\alpha ML/\gamma)[1-(1/2)\log(R_{0}/R)]$ consist of 
the saturation magnetization $M$, the gyromagnetic ratio $\gamma$, the Gilbert damping constant $\alpha$, 
the thickness $L$, the disk radius $R$, and the core radius $R_{0}$ of the free layer. 
The polarity $p$ and chirality $c$ are assumed to be $+1$, for convenience. 
The unit vectors along the $x$ and $z$ axes are denoted as $\hat{\mathbf{x}}$ and $\hat{\mathbf{z}}$, respectively. 
The normalized position of the vortex-core center is $s=|\mathbf{X}|/R$. 
The nonlinear parameter of the damping torque is $\xi(\simeq 1/4)$ \cite{dussaux12}. 
The potential energy is 
\begin{equation}
  W
  =
  \frac{\kappa}{2}
  |\mathbf{X}|^{2}
  +
  \frac{\kappa^{\prime}}{4 R^{2}}
  |\mathbf{X}|^{4},
  \label{eq:potential}
\end{equation}
where $\kappa\simeq (10/9)4\pi M^{2}L^{2}/R$ and $\kappa^{\prime}/\kappa\simeq 1/4$ \cite{guslienko06PRL,dussaux12}. 
The spin-torque strength with the spin polarization $P$ is $a_{J}=\pi \hbar P/(2e)$. 
Throughout the paper, we use the following values of the parameters estimated from experiments and simulations \cite{dussaux12,grimaldi14,tsunegi14,tsunegi16APL} 
as $M=1500$ emu/c.c., $\gamma=1.764\times 10^{7}$ rad/(Oe s), $\alpha=0.005$, $L=4$ nm, $R=150$ nm, 
$R_{0}=10$ nm, and $P=0.3$. 
The direct current $I=\pi R^{2}J$ is $4.0$ mA, corresponding to the current density of $J_{\rm dc} \simeq 5.7 \times 10^{6}$ A/cm${}^{2}$. 
The magnetization in the reference layer is tilted from the $z$ axis with $75^{\circ}$. 
We note that the magnetization direction in the reference layer can be stabilized to the tilted direction in the experiment 
with the help of the interlayer exchange in a synthetic antiferromagnet through a nonmagnetic spacer such as Ru, an exchange bias from an antiferromagnet such as IrMn, 
and an external magnetic field applied to the $z$ direction \cite{grimaldi14,tsunegi14,tsunegi16APL}. 
We note that a finite tilted component $p_{z}$ to the $z$ direction is necessary to excite the dynamics of the magnetic vortex core; see Appendix \ref{sec:AppendixA}. 


\section{Dynamical response to pulse current}
\label{sec:Dynamical response to pulse current}

The input data for the physical reservoir computing are random binary pulse $\mathfrak{b}=0\ {\rm or}\ 1$, 
which is injected as electric current in the spintronics system \cite{torrejon17,markovic19,tsunegi19}. 
In addition, in this work, the output current from the MTJ is returned to the free layer with the delay time $\tau$, 
as schematically shown in Fig. \ref{fig:fig1}(a). 
Therefore, the total current density $J$ in Eq. (\ref{eq:Thiele}) is given by 
\begin{equation}
  J(t)
  =
  J_{\rm dc}
  \left[
    1
    +
    \nu 
    \mathfrak{b}(t)
    +
    \chi 
    \frac{Y(t-\tau)}{R}
  \right],
  \label{eq:current}
\end{equation}
where $\nu$ and $\chi$ are the strengths of the binary pulse and the feedback current, respectively, 
and are set to be $\nu=0.2$ and $\chi=0.1$ in this work. 
The binary pulse is constant during a pulse width $t_{\rm p}$. 
Through the magnetoresistance effect, the magnitude of the feedback current reflects the vortex-core position which has a time lag of delay time $\tau$. 
The resistance of the MTJ with the vortex free layer depends on the $y$ component of the core position \cite{khalsa15} 
because it determines the cross-section area with the magnetic moments parallel (or antiparallel) to $\mathbf{p}$. 
Therefore, the feedback current in Eq. (\ref{eq:current}) is proportional to $Y(t-\tau)$. 
The fourth-order Runge-Kutta method with the continuation method is applied to Eq. (\ref{eq:Thiele}) to include the feedback effect \cite{taniguchi19,hairer93}. 
The reservoir computing is performed by using the output $s(t)$ \cite{torrejon17,tsunegi18}. 



\begin{figure}
\centerline{\includegraphics[width=0.8\columnwidth]{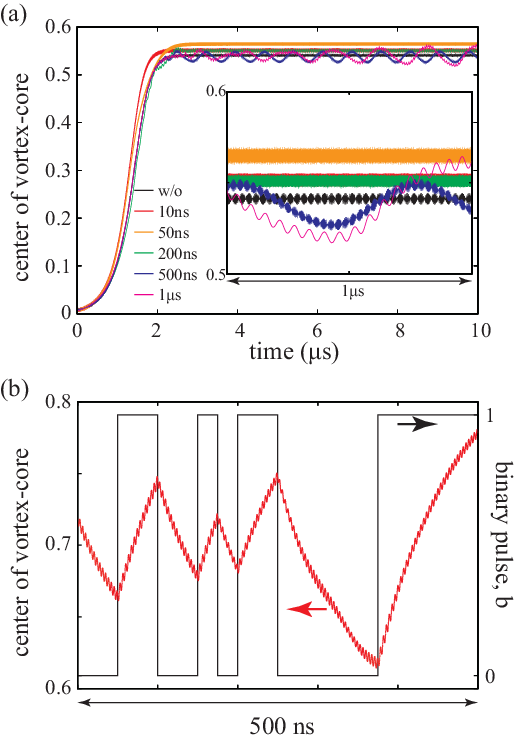}}
\caption{
         (a) Time evolutions of the center of the vortex core with the delay times of 10 ns (red), 50 ns (orange), 200 ns (green), 500 ns (blue), and 1 $\mu$ s (purple).
             The dynamics in the absence of the feedback effect is also shown by the black line. 
             The inset is an enlarged view in a steady state. 
         (b) Dynamics of the center of the vortex core (red) in the presence of random binary pulses (black). 
             The delay time is 500 ns, whereas the pulse width is 25 ns. 
         \vspace{-3ex}}
\label{fig:fig2}
\end{figure}



We note that the output current from the magnetic multilayer depends on the vortex core position through tunnel magnetoresistance effect, as mentioned above. 
Therefore, the memory of the vortex position at a past time is naturally stored in the output signal. 
In other words, by passing through an electric cable, the feedback current naturally carries the information of the past status with a finite delay time. 
Thus, no external registers nor memories devices are necessary, which is preferable feature to use spintronics devices for the physical reservoir computing with the feedback effect. 
Such a delayed-feedback effect without using register nor memory device was already reported experimentally \cite{tsunegi16SR}. 
We note that electric circuits, such as registers and/or memories, can also be applied to the physical reservoir computing because they have memory function and/or nonlinear response. 
It was, however, shown in our previous work that using spintronics device for the reservoir computing results in a large memory functionality compared with an electric circuit without spintronics device \cite{tsunegi18}. 
The result indicates that the spintronics devices contribute to an enhancement of the memory function in the physical reservoir computing. 
Furthermore, we will show in the following that the feedback effect further contributes to achieve a large memory function in the spintronics devices. 


Let us first show the role of the feedback current on the vortex dynamics. 
Figure \ref{fig:fig2}(a) shows the time evolutions of the center position $s(t)$ of the vortex core from the disk center in the absence of random binary pulse. 
The inset shows an enlarged view in a steady state. 
In the absence of the feedback effect ($\chi=0$), the radius of the vortex-core center saturates to a certain value after a relaxation time on the order of 1 $\mu$s 
and shows a limit-cycle oscillation, as indicated by black line. 
A small oscillation of the amplitude is due to the spin-transfer torque originating from the in-plane component ($p_{x}$) of the magnetization in the reference layer, 
which breaks a circular rotation of the vortex core by the torque with the polarization in the $x$ direction. 
The feedback effect changes the vortex dynamics, depending on the delay time. 
For a relatively short delay time such as $10$ (red), $50$ (orange), and $200$ (green) ns, 
the saturated values of the vortex-core position are changed. 
This is due to the modulation of the Hopf bifurcation \cite{khalsa15}, 
where the critical current density necessarily to move the vortex core from the disk center oscillates with the delay time; see also Appendix \ref{sec:AppendixA}. 
An amplitude modulation as well as nonperiodic behavior, is observed for relatively long delay times such as 500 (blue) ns and 1 $\mu$s (purple). 


Figure \ref{fig:fig2}(b) shows a vortex dynamics in the presence of the random binary pulses, where the delay time and the pulse width are $500$ and $25$ ns, respectively. 
The pulse width is sufficiently shorter than the relaxation time of the vortex-core dynamics, 
which is on the order of 1 $\mu$s, as mentioned above. 
Therefore, the dynamics shows a transient behavior from one limit-cycle state to another, and does not saturate to a steady state. 



\section{Reservoir computing and periodic structure in memory capacity}
\label{sec:Reservoir computing and periodic structure in memory capacity}

The procedure of the physical reservoir computing is as follows; 
see also Appendix \ref{sec:AppendixB}. 
The dynamical response $s(t)$ during an input pulse is divided into $N_{\rm node}$ nodes. 
We denote the output $s$ at $i$th node in the presence of $k$th pulse as $s_{k,i}$. 
The time-multiplexing method regards $s_{k,i}$ as the $i$th virtual neuron at a discrete time $k$ \cite{fujii17,torrejon17,tsunegi18}. 
The vortex dynamics can be then regarded as an interacting system consisting of $N_{\rm node}$ neurons, 
where the state of a neuron $s_{k,i}$ is affected by the other neurons through the time evolution described by Eq. (\ref{eq:Thiele}). 
Adding the feedback effect will bring an additional interaction, as discussed below. 
Similarly, let us denote the $k$th binary input data as $\mathfrak{b}_{k}$. 
Applying $N_{\rm L}$ pulse to the vortex, we determine the weight $w_{D,i}$ minimizing the distance between the input and output data, 
\begin{equation}
  \sum_{k=1}^{N_{\rm L}}
  \left(
    \sum_{i=1}^{N_{\rm node}+1}
    w_{D,i}
    s_{k,i}
    -
    \mathfrak{b}_{k-D}
  \right)^{2}. 
\end{equation}
The integer $D(=0,1,2,\cdots)$ is called delay \cite{fujii17}. 
Since the recurrent neural network calculates a time sequence of input data, 
the weight should be defined for each past data injected $D$ times before the current pulse. 
We note that the delay $D$ is a common number used in reservoir computing \cite{fujii17,nakajima19,tsunegi18}, 
and not to confuse with the delay time $\tau$ of the feedback current in the present study. 
Throughout this paper, the symbol $D$ appeared in quantities related to the reservoir computing and used as an integer represents the delay, 
whereas the symbol $D$ appeared in the Thiele equation represents the damping strength.



\begin{figure*}
\centerline{\includegraphics[width=2.0\columnwidth]{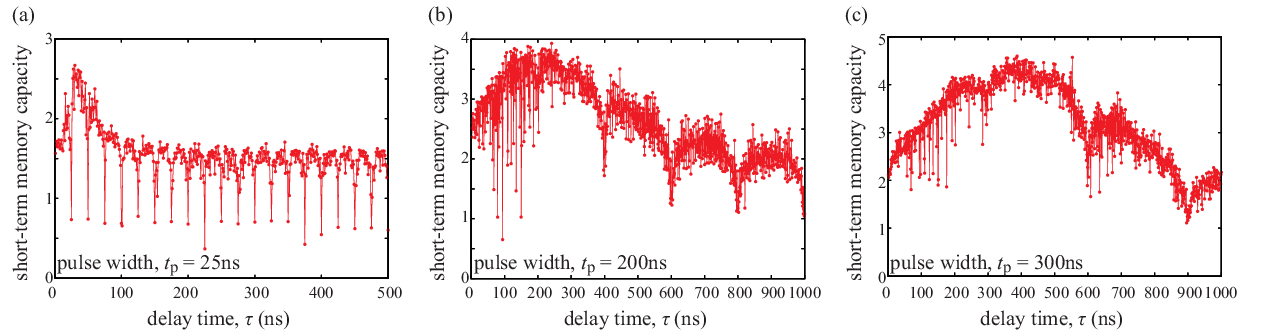}}
\caption{
         Dependences of the short-term memory capacity on the delay time of the feedback current 
         with the pulse width of (a) 25, (b) 200, and (c) 300 ns. 
         \vspace{-3ex}}
\label{fig:fig3}
\end{figure*}




After determining the weights, we injected different random binary pulse $\mathfrak{b}_{m}^{\prime}$, and measure the output $s_{m,i}^{\prime}$, 
where $m=1,2,\cdots,N_{\rm L}^{\prime}$, and the number of the input $N_{\rm L}^{\prime}$ is not necessarily the same with the number $N_{\rm L}$ for the learning. 
Then, we define 
\begin{equation}
  v_{m-D}
  \equiv
  \sum_{i=1}^{N_{\rm node}+1}
  w_{D,i}
  s_{m,i}^{\prime}
\end{equation}
and evaluate the correlation coefficient as 
\begin{equation}
\begin{split}
  &
  {\rm Cor}(D)
  \equiv
\\
  &
  \frac{\sum_{m=1}^{N_{\rm L}^{\prime}} \left(\mathfrak{b}_{m-D}^{\prime} - \langle \mathfrak{b}_{m-D}^{\prime} \rangle \right) \left( v_{m-D} - \langle v_{m-D} \rangle \right)}
    {\sqrt{\sum_{m=1}^{N_{\rm L}^{\prime}}\left(\mathfrak{b}_{m-D}^{\prime} - \langle \mathfrak{b}_{m-D}^{\prime} \rangle \right)^{2}  \sum_{m=1}^{N_{\rm L}^{\prime}}\left( v_{m-D} - \langle v_{m-D} \rangle \right)^{2}}}.
  \label{eq:correlation}
\end{split}
\end{equation}
The correlation coefficient characterizes the reproducibility of the input data $\mathfrak{b}_{m}^{\prime}$ from the output $s_{m,i}^{\prime}$, 
and quantifies the memory function of the vortex free layer. 
When the input data can be reproduced completely, $[{\rm Cor}(D)]^{2} \to 1$, 
whereas $[{\rm Cor}(D)]^{2} \to 0$ when the vortex free layer cannot reproduce the input data. 
The short-term memory is defined as 
\begin{equation}
  C_{\rm STM}
  \equiv
  \sum_{D=1}^{\infty}
  \left[
    {\rm Cor}(D)
  \right]^{2}.
  \label{eq:C_STM}
\end{equation}




\begin{figure*}
\centerline{\includegraphics[width=2.0\columnwidth]{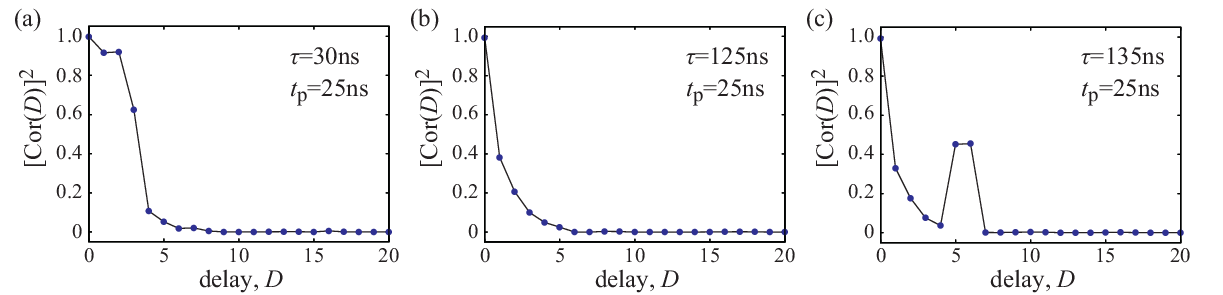}}
\caption{
         The square of the correlation coefficient, $[{\rm Cor}(D)]^{2}$, as a function of the delay $D$ for the delay time $\tau$ of 
         (a) 30, (b) 125, and (c) 130 ns, where the pulse width is 25 ns. 
         \vspace{-3ex}}
\label{fig:fig4}
\end{figure*}




Figures \ref{fig:fig3}(a)-\ref{fig:fig3}(c) show the short-term memory capacity, as a function of the delay time, for the pulse widths of 25, 200, and 300 ns. 
A large short-term memory capacity due to the presence of the feedback effect is found, in particular for the delay time slightly longer than the pulse width; 
see Appendix \ref{sec:AppendixC} for comparison, where the short-term memory capacity in the absence of the feedback effect is shown. 
We also show the value of the short-term memory capacity in echo-state network \cite{jaeger04} in Appendix \ref{sec:AppendixD} for comparison. 
An important finding in this work is a periodic behavior of the short-term memory capacity with respect to the delay time; 
i.e., the short-term memory shows sharp drops at the delay time of $\tau \simeq n t_{\rm p}$, where $n$ is an integer. 
The results reveal that the feedback effect does not always contribute to the enhancement of the memory function. 



\begin{figure}
\centerline{\includegraphics[width=1.0\columnwidth]{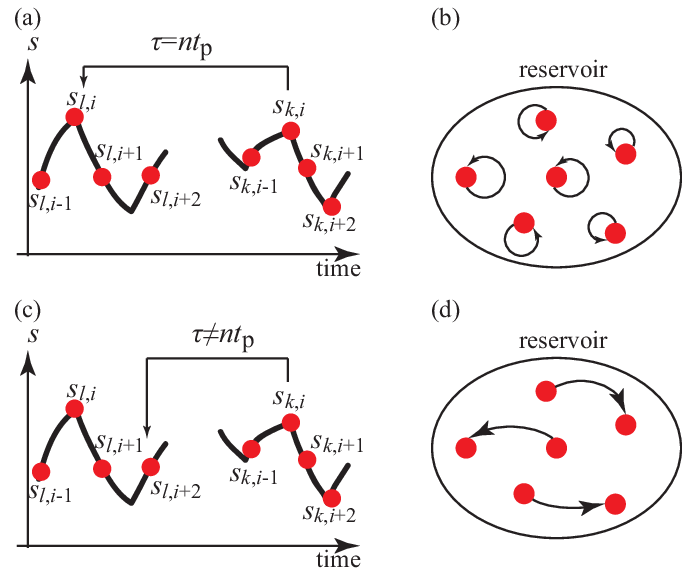}}
\caption{
         (a) The time evolution of the $i$th virtual node at time $k$ depends on itself in past time $l$ via the feedback effect when the delay time is an integer multiple of the pulse width. 
         (b) The feedback effect can be regarded as a self-interaction in this case. 
         Those ones when the delay time is not an integer multiple of the pulse width are shown in panels (c) and (d). 
         \vspace{-3ex}}
\label{fig:fig5}
\end{figure}




The periodic behavior of the short-term memory capacity can also be implied from the correlation coefficients. 
Note that the correlation coefficients in the absence of the feedback effect monotonically decrease with increasing the delay $D$; see Ref. \cite{tsunegi18} and Appendix \ref{sec:AppendixC}. 
On the other hand, Figs. \ref{fig:fig4}(a)-\ref{fig:fig4}(c) show the correlation coefficients in the presence of the feedback effect 
for $\tau=30$, $125$, and $135$ ns, respectively, where the pulse width is fixed to 25 ns. 
The correlation coefficients show non-monotonic behavior and has a local maximum at the delay $D$ near $D\simeq \tau/t_{\rm p}$ 
when the delay time is not an integer multiple of the pulse width, as shown in Figs. \ref{fig:fig4}(a) and \ref{fig:fig4}(c). 
The local maximum of the correlation coefficients contributes to the large short-term memory capacity. 
The monotonic decrease of the correlation coefficients, on the other hand, appears when the delay time is an integer multiple of the pulse width, as shown in Fig. \ref{fig:fig4}(b). 
The short-term memory capacity becomes small in this case. 




We consider the origin of the periodic behavior of the short-term memory capacity, 
although the relation between the network structure and memory function in the reservoir computing is not fully clarified yet \cite{rodan10,kawai19,rodriguez19}. 
We note that the feedback effect in Eq. (\ref{eq:Thiele}) can be regarded as an interaction between virtual neurons. 
The presence of the feedback current in Eq. (\ref{eq:Thiele}) means that the time evolution of a virtual neuron $s_{k,i}$ depends on other neuron $s_{l,j}$ ($l<k$) in the past \cite{comment1}. 
When the delay time is an integral multiple of the pulse width ($\tau=n t_{\rm p}$), the neuron in the delayed information is the present neuron itself, i.e., $j=i$; see Fig. \ref{fig:fig5}(a). 
Therefore, the feedback effect in this case is a self-feedback, and does not provide any interaction between different neurons, as schematically shown in Fig. \ref{fig:fig5}(b). 
On the other hand, when $\tau \neq n t_{\rm p}$, the information of a different neuron $s_{l,j}$ ($j \neq i$) is used to determine the time evolution of the present neuron, $s_{k,i}$; see Fig. \ref{fig:fig5}(c). 
The feedback effect in this case can be regarded as an interaction between different neurons, as shown in Fig. \ref{fig:fig5}(d). 
It is useful to remember that the memory function of the physical reservoir comes from the recurrent network between neurons \cite{mandic01,maas02,jaeger04,verstraeten07,appeltant11}. 
Since the feedback effect does not provide interactions between the neurons for $\tau = n t_{\rm p}$, the memory function of the physical reservoir remains small. 
On the other hand, the feedback effect provides an additional interaction between the neurons for $\tau \neq n t_{\rm p}$, resulting in the enhancement of the memory function. 
As a result, a periodic structure of the short-term memory capacity with respect to the delay time appears with the interval of $n t_{\rm p}$.

\section{Conclusion}
\label{sec:Conclusion}

In conclusion, the physical reservoir computing by using a vortex-type ferromagnet in the presence of the feedback current is performed theoretically. 
A periodic behavior of the short-term memory capacity was found with respect to the delay time of the feedback current, 
where the memory capacity becomes large when the delay time $\tau$ is not an integer multiple of the pulse width $t_{\rm p}$, 
whereas it shows sharp drops at the delay time $\tau$ satisfying $\tau=n t_{\rm p}$ with a positive integer $n$. 
The origin of the periodic structure can be attributed to be an interaction between the virtual neurons via the feedback current.


\section*{Acknowledgement}

The authors acknowledge Shinji Miwa for valuable discussions. 
The results were partially obtained from a project [Innovative AI Chips and Next-Generation Computing Technology Development/(2) 
Development of next-generation computing technologies/Exploration of Neuromorphic Dynamics towards Future Symbiotic Society] commissioned by NEDO. 
K. N. is supported by 
JSPS KAKENHI Grant No. JP18H05472.


\appendix


\section{Hopf-bifurcation analysis}
\label{sec:AppendixA}

The Thiele equation, Eq. (\ref{eq:Thiele}), in terms of the normalized core position $s$ and the phase $\psi$ from the $x$ axis is given by 
\begin{widetext}
\begin{equation}
  \dot{s}
  =
  \frac{Ga_{J}Jp_{z}-|D|\kappa (1+\xi s^{2})(1+\zeta s^{2})}{G^{2}+(1+\xi s^{2})^{2}D^{2}}
  s
  +
  \frac{ca_{J}Jp_{x}R_{0}/R}{G^{2}+(1+\xi s^{2})^{2}D^{2}}
  \left[
    |D| (1 + \xi s^{2}) \cos\psi
    -
    G \sin \psi
  \right],
\end{equation}
\begin{equation}
  \dot{\psi}
  =
  \frac{G \kappa (1+\zeta s^{2}) + |D| (1+\xi s^{2})a_{J}Jp_{z}}{G^{2}+(1+\xi s^{2})^{2}D^{2}}
  -
  \frac{c a_{J} J p_{x} R_{0}/R}{G^{2}+(1+\zeta s^{2})^{2} D^{2}}
  \left[
    |D| (1 + \xi s^{2}) \sin\psi
    +
    G \cos \psi
  \right]
  \frac{1}{s}.
\end{equation}
\end{widetext}

We note that the vortex-core size, $R_{0}$, is usually much smaller than the disk radius; see Sec. \ref{sec:System description} for the values of the parameters. 
In addition, the Gilbert damping constant $\alpha$ in conventional ferromagnets, such as Fe, Co, Ni, and their alloys, are usually small ($\alpha\ll 1$) \cite{oogane06}. 
Therefore, neglecting the spin-transfer torque, proportional to $R_{0}/R$, originated from an in-plane component of the magnetization $\mathbf{p}$ 
and higher order terms of $|D| \propto \alpha$ and $s(<1)$, 
the equation of motion of $s$ becomes the Stuart-Landau equation for the real variable $s$ as 
\begin{equation}
  \dot{s}
  =
  as
  -
  bs^{3}, 
\end{equation}
where $a$ and $b$ are defined as 
\begin{align}
  a
  =
  \frac{|D| \kappa}{G^{2}}
  \left(
    \frac{Ga_{J}Jp_{z}}{|D|\kappa}
    -
    1
  \right),
&&
  b
  =
  \frac{|D| \kappa}{G^{2}}
  \left(
    \xi 
    +
    \zeta 
  \right).
\end{align}
The solution of the Stuart-Landau equation with the initial condition of $s(0)$ is 
\begin{equation}
  s(t)
  =
  \frac{s(0) e^{at}}{\sqrt{1+ s(0)^{2}(b/a)(e^{2at}-1)}}.
\end{equation}
The solution indicates that the Hopf bifurcation appears at the current satisfying $a=0$, 
i.e., the vortex core is stabilized at the disk center ($s=0$) when $J<J_{\rm c1}$, whereas the vortex core moves from the disk center to a limit cycle oscillation with the radius $\lim_{t\to \infty}s(t)=\sqrt{a/b}$ when $J>J_{\rm c1}$, 
where the critical current density $J_{\rm c1}$ is given by 
\begin{equation}
  J_{\rm c1}
  =
  \frac{|D|\kappa}{Ga_{J}p_{z}}. 
\end{equation}
We note that a finite tilted component of the magnetization in the reference layer, $p_{z}$, is necessary 
to move the vortex core from the disk center. 
This is because an injection of energy to the system becomes finite by the work done by the spin-transfer torque originating from the perpendicular component of $\mathbf{p}$. 
We note that the above analysis works well to explain the limit cycle oscillation of the vortex core. 
For example, using the values of the parameters mentioned in Sec. \ref{sec:System description}, 
$\lim_{t\to \infty}\sqrt{a/b}=0.55$, which agrees with the saturated value of $s$ in the absence of the feedback effect shown in Fig. \ref{fig:fig2}(a). 

We also note that the normalized radius should satisfy $s \le 1$. 
Therefore, the current density should be smaller than $J_{\rm c2}=J_{\rm c1}(1+\xi+\zeta)$. 
In summary, the Hopf bifurcation appears at the current density of $J_{\rm c1}$, 
and the limit cycle oscillation is excited when the current density is in the range of $J_{\rm c1}<J<J_{\rm c2}$.


\section{Evaluation method of short-term memory capacity}
\label{sec:AppendixB}

Here, we explain the procedure of the physical reservoir computing by focusing on the evaluation method of the short-term memory capacity \cite{fujii17,furuta18,tsunegi18}. 
In general, a physical system consisting of $N$ interacting particles is considered as a reservoir, where $N$ is the number of the particles. 
In this section, however, we consider a particle, and regard its dynamical response to input data as an ensemble of numerous virtual neurons, as done in experiments. 

\begin{figure}
\centerline{\includegraphics[width=1.0\columnwidth]{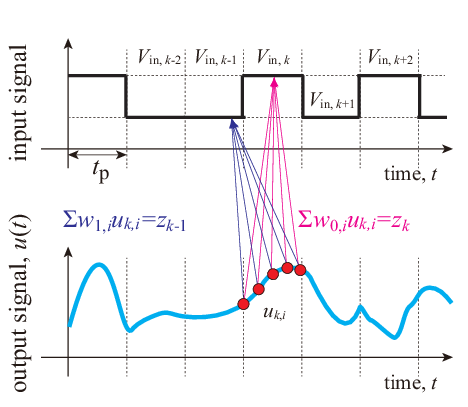}}
\caption{
         Schematic view of the learning process. 
         The weight $w_{D,i}$ is determined to reproduce the target output $z_{k-D}$ from the system output $u_{k,i}$. 
         The figure shows the case evaluating the short-term memory capacity, 
         where the target output $z_{k}$ is the input signal $V_{{\rm in},k}$ (or equivalently, the random binary pulse $\mathfrak{b}_{k}$). 
         }
\label{fig:fig6}
\end{figure}

We first apply $N_{\rm L}$-bit random input into the system, as shown in Fig. \ref{fig:fig6}. 
As an example, let us assume that the input signal is an electric voltage $V$, as done in experiments \cite{tsunegi18,torrejon17}. 
The input data consists of a bias term $V_{\rm bias}$ and random binary data $V_{\rm p}$ as 
\begin{equation}
  V_{{\rm in},k}
  =
  V_{\rm bias}
  +
  V_{\rm p}
  \mathfrak{b}_{k},
  \label{eq:input}
\end{equation}
where $\mathfrak{b}_{k}(=0\ {\rm or}\ 1)$ represents the $k$th binary pulse. 
Note that the binary pulse, $V_{\rm p}\mathfrak{b}_{k}$, is constant during a pulse width $t_{\rm p}$. 
The $k$th binary pulse is applied to the system in the time range of $(k-1) t_{\rm p} \le t < kt_{\rm p}$. 
The number of the input is $N_{\rm L}$. 




\begin{figure}
\centerline{\includegraphics[width=1.0\columnwidth]{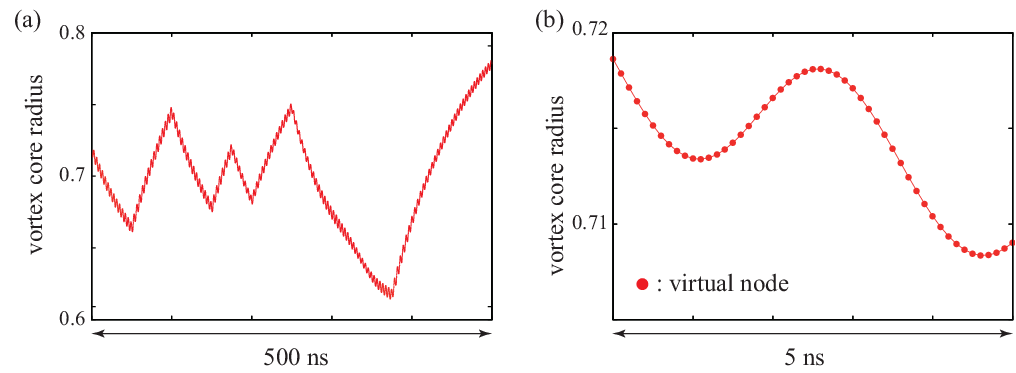}}
\caption{
         (a) Dynamics of the vortex-core radius $s$ in the presence of random binary pulses, 
         and (b) the definition of the virtual nodes from the vortex-core radius. 
         The pulse width is 25 ns, whereas the node number is $N_{\rm node}=250$. 
         }
\label{fig:fig7}
\end{figure}

The physical system shows some output, such as an oscillating voltage emitted from spin-torque oscillator \cite{torrejon17}. 
In experiments \cite{tsunegi18,torrejon17,tsunegi19}, the output voltage is divided into the amplitude and phase by the Hilbert transformation. 
The amplitude was used for the reservoir computing in Refs. \cite{tsunegi18,torrejon17}, whereas Ref. \cite{tsunegi19} used the phase for the computation. 
Although the main text in this work uses the (normalized) amplitude $s=|\mathbf{X}|/R$ of the vortex core for the reservoir computing, 
the following procedure is applicable to the computation using the phase. 
Let us denote therefore a general output from the system as $u(t)$. 
As explained in the main text, we divide the output $u(t)$ in the presence of the $k$th input $V_{{\rm in},k}$ into $N_{\rm node}$ nodes, 
and denote it as $u_{k,i}$, as shown in Fig. \ref{fig:fig6}, i.e., 
\begin{equation}
  u_{k,i}
  =
  u 
  \left(
    \left(
      k - 1 
    \right)
    t_{\rm p}
    +
    i 
    \frac{t_{\rm p}}{N_{\rm node}}
  \right).
\end{equation}
The discrete output $u_{k,i}$ is regarded as the $i$th neuron at time $k$. 
Figure \ref{fig:fig7} shows how the virtual nodes are defined from the magnetic vortex-core dynamics studied in the main text. 
The dynamical response of the vortex-core radius $s(t)$ in the presence of the random binary pulses is shown in Fig. \ref{fig:fig7}(a), 
which is the same with that shown in Fig. \ref{fig:fig2}(b) in the main text. 
Figure \ref{fig:fig7}(b) is an enlarged view during a relatively short time range. 
In this case, the pulse width $t_{\rm p}$ is 25 ns, whereas the node number $N_{\rm node}$ is 250. 
Therefore, the virtual nodes appear with the time interval of 0.1 ns, as indicated by red circles in Fig. \ref{fig:fig7}(b). 



The physical reservoir computing aims to reproduce some of the data from the output $u_{k,i}$. 
The data to be reproduced is not necessarily the same with the input data, $V_{{\rm in},k}$, or equivalently, $\mathfrak{b}_{k}$. 
In general, the data to be estimated from the output $u_{k,i}$ is called target data. 
Let us denote the target data as $z_{k}$, in general. 
For the evaluation of the short-term memory capacity, 
the target data is the random binary pulse, i.e., 
\begin{equation}
  z_{k}
  =
  \mathfrak{b}_{k}. 
\end{equation}
On the other hand, for the evaluation of the parity check capacity, discussed in Refs. \cite{furuta18,tsunegi19}, 
the target output is chosen as 
\begin{equation}
  z_{k-D}
  =
  \sum_{j=0}^{D}
  \mathfrak{b}_{k-D+j}\ \ {\rm mod}\ 2.
\end{equation}


Weight $w_{D,i}$ is introduced to find one-to-one correspondence between the system output $u_{k,i}$ and the target output $z_{k}$. 
The weight is determined to minimize the error between the system output and the target output given by 
\begin{equation}
  \sum_{k=1}^{N_{\rm L}}
  \left(
    \sum_{i=1}^{N_{\rm node}+1}
    w_{D,i}
    u_{k,i}
    -
    z_{k-D}
  \right)^{2}.
\end{equation}
The bias term of the system output is $u_{k,N_{\rm node}+1}=1$ \cite{fujii17}. 
The weight $w_{D,i}$ is set so as to reproduce the $(k-D)$th target output from the system output $u_{k}$ at time $k$. 
Note that the weight should be introduced for each kind of the target output, 
i.e., the weight for the evaluation of the short-term memory capacity is different from that for the evaluation of the parity check capacity. 
The process determining the weight is called learning, 
and the target output $z_{k}$ for the learning is called training data. 


After the learning, other random binary data are injected into the system, where the number of the input data $N_{\rm L}^{\prime}$ is not necessarily the same with $N_{\rm L}$ for the learning. 
We denote the system and target output in this process as $u_{n,i}^{\prime}$ and $z_{n}^{\prime}$ to distinguish them from $u_{k,i}$ and $z_{k}$ for the learning. 
Using the system output $u_{n,i}^{\prime}$ with the weight determined by the learning, we define 
\begin{equation}
  v_{n-D}
  \equiv
  \sum_{i=1}^{N_{\rm node}+1}
  w_{D,i}
  u_{n,i}^{\prime}. 
\end{equation}
If the learning is sufficiently achieved, $v_{n-D}$ will reproduce the target output well. 
To quantify the reproducibility, we evaluate the correlation coefficient given by 
\begin{widetext}
\begin{equation}
  {\rm Cor}(D)
  \equiv
  \frac{\sum_{n=1}^{N_{\rm L}^{\prime}} \left(z_{n-D} - \langle z_{n-D} \rangle \right) \left( v_{n-D} - \langle v_{n-D} \rangle \right)}
    {\sqrt{\sum_{n=1}^{N_{\rm L}^{\prime}}\left(z_{n-D} - \langle z_{n-D} \rangle \right)^{2}  \sum_{n=1}^{N_{\rm L}^{\prime}}\left( v_{n-D} - \langle v_{n-D} \rangle \right)^{2}}}. 
  \label{eq:correlation_general}
\end{equation}
\end{widetext}
In the main text, $u_{k,i}$, as well as $u_{n,i}^{\prime}$, and $z_{n}$ are the normalized vortex-core radius $s_{k,i}$ and the random binary $\mathfrak{b}_{n}$ 
because we focus on the evaluation of the short-term memory capacity. 
If we use, on the other hand, the target output for the parity check, the correlation coefficient for the evaluation of the parity check capacity will be evaluated. 
The capacity is generally defined as 
\begin{equation}
  C
  \equiv
  \sum_{D=1}^{\infty}
  \left[
    {\rm Cor}(D)
  \right]^{2}.
\end{equation}
Note that the delay $D$ is defined from zero ($D=0,1,2,\cdots$), 
whereas the capacity in this work is defined from the correlation coefficients for $D \ge 1$, as done in previous works \cite{fujii17,furuta18,tsunegi18,tsunegi19}. 
The maximum value of the capacity is the number of nodes, i.e., $C \le N_{\rm node}$. 


In the present work, we first solve the Thiele equation without random binary pulse from $t=0$ to $t=5$ $\mu$s. 
After that, the random pulses are applied as washout, and then, $N_{\rm L}=1000$ pulses are applied for the learning, 
where the number of the pulses used for the washout is 300. 
Then, different 300 random pulses for the washout are applied again. 
After that $N_{\rm L}^{\prime}=1000$ pulses are applied to evaluate the short-term memory capacity. 




\begin{figure}
\centerline{\includegraphics[width=1.0\columnwidth]{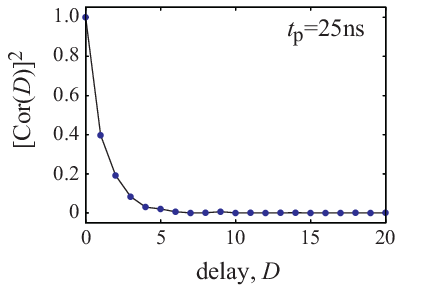}}
\caption{
        Square of the correlation coefficient in the absence of the feedback effect. 
        The pulse width is 25 ns. 
         }
\label{fig:fig8}
\end{figure}



\section{Short-term memory capacity in the absence of feedback effect}
\label{sec:AppendixC}

In the main text, we show the correlation coefficients and the short-term memory capacity in the presence of the feedback current. 
For comparison, here, we show these quantities in the absence of the feedback effect. 

Figure \ref{fig:fig8} shows the square of the correlation coefficients obtained from the Thiele equation without feedback effect, 
where the pulse width is $t_{\rm p}=25$ ns. 
The correlation decreases monotonically and rapidly with the increase of the delay $D$, as found in experiments \cite{tsunegi18,tsunegi19}. 
We remind the readers that a similar behavior is observed even in the presence of the feedback effect when the delay time $\tau$ of the feedback current satisfies $\tau\simeq n t_{\rm p}$ ($n \in \mathbb{N}$). 
The short-term memory capacity evaluated from Fig. \ref{fig:fig8} is 0.74, which is smaller than the value with the feedback effect satisfying $\tau \neq n t_{\rm p}$; see Fig. \ref{fig:fig3}.


\section{Comparison with echo-state network}
\label{sec:AppendixD}



\begin{figure}
\centerline{\includegraphics[width=1.0\columnwidth]{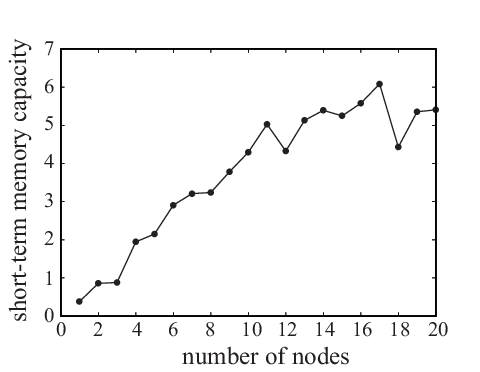}}
\caption{
         Short-term memory capacity of echo-state network as a function of the number of nodes. 
         }
\label{fig:fig9}
\end{figure}



Here, we evaluate the short-term memory capacity of echo-state network \cite{jaeger04,fujii17,furuta18} and compare it with the spintronics reservoir with the feedback effect. 
Echo-state network consists of $N_{\rm node}$-nodes given as $\mathbf{x}(t_{\rm d})=[x_{1}(t_{\rm d}),x_{2}(t_{\rm d}),\cdots,x_{N_{\rm node}}(t_{\rm d})]$, 
where $t_{\rm d}=1,2,\cdots$ represents a discrete time. 
The time evolution of the nodes is given by 
\begin{equation}
  x_{i}(t_{\rm d}+1)
  =
  \tanh
  \left[
    \sum_{j=1}^{N_{\rm node}}
    W_{ij}
    x_{j}(t_{\rm d})
    +
    W_{{\rm in},i}
    \mathfrak{b}(t_{\rm d}+1)
  \right],
  \label{eq:ESN}
\end{equation}
where $\mathfrak{b}$ is the binary input. 
The components of $N_{\rm node}$th order matrix $\mathsf{W}$ and vector $\mathbf{W}_{\rm in}$, $W_{ij}$ and $W_{{\rm in},i}$, are random numbers. 
The matrix $\mathsf{W}$ corresponds to the weight of the internal connection between nodes, whereas the vector $\mathbf{W}_{\rm in}$ corresponds to the weight to the input data \cite{fujii17}. 
In this work, we normalized $\mathbf{W}_{\rm in}$ by dividing its component by the absolute value of the component, $|{\rm max}[W_{{\rm in},i}]|$. 
In addition, we evaluate the singular value of $\mathsf{W}$, and divide the components of $\mathsf{W}$ by the maximum singular value. 
The maximum singular value of the internal weight $\mathsf{W}$ is called spectral radius \cite{fujii17}. 
The normalization of $\mathsf{W}$ makes the spectral radius of the present calculation 1.0. 
A training noise is neglected, for simplicity. 
Similar calculations are reported in Refs. \cite{fujii17,furuta18}. 


We emphasize the difference between the spintronics reservoir in the main text and the echo-state network. 
The spintronics device in the main text consists of single device, and the virtual neurons are defined by dividing the dynamical response during a pulse input into $N_{\rm node}$-nodes. 
On the other hand, the echo-state network consists of $N_{\rm node}$-nodes, and we do not divide the time evolution of $\mathbf{x}$ during a pulse.

The system output $u_{k,i}$ defined in Appendix \ref{sec:AppendixB} corresponds to $x_{i}(k)$ of the echo-state network. 
Applying the procedure explained in Appendix \ref{sec:AppendixB}, the short-term memory capacity of the echo-state network can be evaluated. 
In this work, we use the same numbers of the training data and washout with those used in the spintronics reservoir. 
Figure \ref{fig:fig9} shows the dependence of the short-term memory capacity of the echo-state network on the node numbers, $N_{\rm node}$. 
Although the value of the short-term memory capacity depends on the random numbers in $\mathsf{W}$ and $\mathbf{W}_{\rm in}$, as well as the spectral radius and the training noise, 
and thus, the values in Fig. \ref{fig:fig9} might not be an optimized, 
the values are close to those found in previous works \cite{fujii17,furuta18}. 
For example, the short-term memory capacity for $N_{\rm node}=5$ is about $2.2$, which is smaller than the maximum values of the short-term memory capacities of the spintronics reservoir with the feedback effect; 
see Figs. \ref{fig:fig3}(a)-\ref{fig:fig3}(c). 
The result in Fig. \ref{fig:fig9} implies that a single spintronics reservoir with the feedback current is comparable to an echo-state network consisting of 5-10 nodes. 



%


\begin{thebibliography}{49}
\bibitem [{com()}]{comment1}%
  \BibitemOpen
  \href@noop {} {}\bibinfo {note} {There is another possibility that a virtual
  node is absent at time $\tau$ before the present time, depending on the
  values of $t_{\rm p}$, $N_{\rm node}$, and $\tau$. We exclude this case here
  because the number of nodes ($N_{\rm node}=250$) is sufficiently large, and
  therefore, the delay time $\tau$ is always an integral multiple with the
  interval of the nodes given by $t_{\rm p}/N_{\rm node}$.}\BibitemShut {Stop}%
\end{thebibliography}


\end{document}